\documentclass[preprint,preprintnumbers,amsmath,amssymb]{revtex4}

\usepackage{graphicx}
\usepackage{dcolumn}
\usepackage{bm}

\begin{document}


\title{\bf Structure and dielectric response in the high $T_c$ ferroelectric Bi(Zn,Ti)O$_3$-PbTiO$_3$ solid solutions}

\author{ Ilya Grinberg$^1$, Matthew R. Suchomel$^{2}$, Wojtek Dmowski$^3$, Sara E. Mason$^1$, Hui Wu$^2$, Peter K. Davies$^2$ and Andrew M. Rappe$^1$}

\address{$^1$ The Makineni Theoretical Laboratories, Dept. of Chemistry, University of Pennsylvania, Philadelphia, PA 19104-6323\\
$^2$ Dept. of Materials Science and Engineering, University of Pennsylvania, Philadelphia, PA 19104-6323\\
 $^3$ Dept. of Materials Science and Engineering, University    of Tennessee, Knoxville, TN 37996\\
}

\date{\today}

\begin{abstract}
Theoretical {\em ab initio} and experimental methods were used to investigate the 
$\left[{\rm Bi}\left({\rm Zn}_{1/2}{\rm Ti}_{1/2}\right){\rm O}_3\right]_x \left[{\rm Pb}{\rm Ti}{\rm O}_3\right]_{1-x}$
(BZT-PT) solid solution.   We find that hybridization between Zn 4$p$ and O 2$p$ orbitals allows the formation of short, covalent Zn-O bonds, enabling favorable coupling between A-site and B-site displacements.  This leads to large polarization, strong tetragonality and an elevated ferroelectric to paraelectric phase transition temperature. Inhomogeneities in local structure near the 90$^\circ$ domain boundaries  can be deduced from the asymetric peak broadening in the neutron and x-ray diffraction spectra.  These extrinsic effects make the ferroelectric to paraelectric phase transition diffuse in BZT-PT solid solutions.
\end{abstract}

\maketitle

Due to their importance in ferroelectric and piezoelectric applications and their fascinating physical properties, perovskite solid solutions have been an area of intense research~\cite{Cohen92p136,Bellaiche00p5427,Grinberg02p909,Saito04p84,Eitel01p5999}. Recently, it was shown that very large $c/a$ ratios (1.11) and high $T_c$ ($\approx$1000~K) can be obtained in the 
$\left[{\rm Bi}\left({\rm Zn}_{1/2}{\rm Ti}_{1/2}\right){\rm O}_3\right]_x \left[{\rm Pb}{\rm Ti}{\rm O}_3\right]_{1-x}$  (BZT-PT) solid solution~\cite{Suchomel05p262905}. This makes BZT-PT a promising candidate material for use in next-generation piezoelectric and nonvolatile memory applications. In particular, the rate of increase in both properties as BZT replaces PT is the largest known for any PT-based perovskite solid solution. Experimental investigations of dielectric properties in the BZT-PT system also show that the ferroelectric to paraelectric phase transition becomes increasingly diffuse with greater BZT concentration.
This is unusual, as increased tetragonality and stronger ferroelectricity is typically accompanied by a sharpening of the dielectric constant peak at the $T_c$ and loss of relaxor and/or diffuse phase transition behavior~\cite{Landolt}. 
We note that the Bi(Mg,Ti)O$_3$ - PT (BMT-PT) solid solution, which does not exhibit such anomalous behavior~\cite{Suchomel04p4405}, differs from the BZT-PT system only by the isovalent B site cation replacement of Mg for Zn.  
In this Letter, we combine theoretical calculations with new experimental data to reveal the origins of anomalous behavior in the BZT-PT system.

To determine the local, \AA-scale structure in the BZT-PT system,  we performed density functional theory~\cite{Perdew81p5048}  (DFT) calculations using $2 \times 2 \times 2$ 40-atom  supercells with periodic boundary conditions at experimental volume~\cite{Suchomel05p262905}.   Technical details are the same as in previous work~\cite{Grinberg04p220101}.
We use five and seven different cation arrangements with minimal oxygen over- and under-bonding~\cite{Cockayne99pR12542,Grinberg04p144118, Grinberg04p220101} to study $x$=0.25 and $x$=0.5 compositions respectively.  Polarization magnitudes were calculated using the Berry-phase approach~\cite{Kingsmith93p1651}.  

Results of theoretical calculations were compared with experimental structural information obtained from neutron powder diffraction (NPD) of selected compositions in the ($x$)BZT - (1-$x$)PT system. The neutron powder diffraction data were collected under ambient conditions over the range 3-165$^\circ$ 2$\theta$ using BT-1 32 detector neutron powder diffractometer at the NIST Center for Neutron Research.  Full details of the experimental data and the Rietveld structural refinements are available as supplementary online material. Additional structural information was extracted from the neutron scattering data collected at NPDF spectrometer at LANSCE, Los Alamos National Lab by pair distribution function (PDF) analysis techniques; details of this experimental procedure have been described previously~\cite{Juhas04p214101}.

A summary of structural data obtained from both DFT calculations and refined NPD data are shown in Table I.  This compares the measured and predicted (100) cation displacements for several compositions in the ($x$)BZT - (1-$x$)PT system and reveals very good agreement between the theoretically and experimentally determined values.  While displacements determined by refinement of the NPD data show only average A and B site behavior, in the DFT calculations it is possible to distinguish between the chemically distinct cations occupying each cation site in the tetragonal perovskite structure.  A closer examination of the different behavior of the cations in the ($x$)BZT - (1-$x$)PT system helps us understand its anomalous behavior.

We find that Zn distortions are very large, reaching 0.5~\AA\ in the  $x$=0.50 composition.  The combination of the low valence of Zn and the long, $\approx$2.6~\AA\  Zn-O distances created by the large displacements eliminates the bond between Zn and one of its six O neighbors, transforming the  Zn-O$_6$ cages are into Zn-O$_5$ complexes.  
The unusually large Zn distortions are due to the hybridization of 4$s$ and 4$p$ orbitals of Zn with 2$p$ orbitals of O atoms.   Local atom projected density of states (LDOS) (Figure 1a) shows that Zn 4$s$ and 4$p$ orbitals are partially filled in ($x$)BZT-(1-$x$)PT, due to covalent bonding with oxygen 2$p$ states.  This is similar to the LDOS of covalently bonded Ti ions in PbTiO$_3$ (where the hybridization is between Ti 3$d$ and O 2$p$)~\cite{Cohen92p136}, and is in stark contrast to the highly ionic bonding of Mg ions in BMT-PT~\footnote[1]{Calculated LDOS for BMT-PT shows that in contrast to Zn, the Mg 4$s$ states are completely empty.}. 
  In comparison with Zn,  Ti displacements in the ($x$)BZT - (1-$x$)PT system do not change with alloying $x$.  From $x$=0.25 to $x$=0.50, the calculated Ti cation displacement only increases from 0.32~\AA\ to 0.34~\AA.  Even for $x$=0.50 composition, the Ti-O$_6$ cage is largely preserved, with average bond order $\approx$0.25~\cite{Brown81p1} for the long Ti-O bond distances.  The large Ti displacements are enabled by the Zn off-centering, due to coupling between B-cation displacements~\cite{Ghosez99p836,Grinberg04p220101}.

Both our experimental and theoretical data (Table I) show that large Bi and Pb distortions are present  in the BZT-PT solid solutions.  The typical Bi distortion magnitudes are about 0.80~\AA\ and 0.90~\AA\ for $x$=0.25 and $x$=0.50, respectively.    The magnitudes of Bi displacements found in this study are much larger than the 0.5~\AA\ distortion found for Bi in BiFeO$_3$~\cite{Neaton05p014113}. Addition of BZT also induces a 10-20$\%$ increase in Pb displacement magnitudes.   
Analysis of Bi-O$_{12}$ and Pb-O$_{12}$ cages  reveals a crucial difference between the bonding behavior of the two A-site cations.
   Bi displacements create three or four very short ($\approx$2.1-2.2 \AA) and strong (bond order $\approx$0.6-0.8) Bi-O bonds. This motif was also found by a recent  x-ray absorption fine structure study of  the (K,Bi)TiO$_3$-(Na,Bi)TiO$_3$ solid solution~\cite{Shuvaeva05p174114}.  
Bond valence analysis~\cite{Brown81p1} shows that typically 70-90$\%$  of the Bi-O  bonding is with these three or four closest O atoms, while the other O atoms are located far enough from Bi to make their contribution to the bonding negligible.  Essentially, the large Bi displacements create a Bi-O$_3$ or Bi-O$_4$ complex out of the Bi-O$_{12}$ cage.  In contrast, although the Pb atoms also make three or four short Pb-O bonds, the larger ionic size of Pb and its smaller displacement magnitude mean that  the differences among Pb-O bond lengths are smaller. The shortest three or four Pb-O bonds typically account account for only 40-60$\%$ of the total Pb-O bond order. Thus, the  difference in ionic sizes  of Pb$^{2+}$ (1.49\AA) and Bi$^{3+}$ (1.36\AA) gives rise to different A-O bonding motifs.

In agreement with DFT results, we find that substitution of  BZT for PT leads to a broad pair distribution function (PDF), indicative of a material with strong local distortions from overall structure.  Comparison of the neutron-scattering PDF for $x$=0.20 and the relaxed DFT PDF for $x$=0.25 shows excellent agreement between theory and experiment (Figure 1b).  The shoulder at 2.2~\AA\ in the experimental PDF is due to the short Bi-O distances created by the large Bi displacements. In the higher-resolution DFT PDF, another shoulder is apparent at 2.5~\AA, consistent with the off-center displacement of Pb ions~\cite{Egami98p1,Grinberg04p144118}.   



The local structure distortions determined by our DFT calculations manifest themselves in the observed structural and dielectric properties of the BZT-PT solid solution. The increase in the $c$ axis lattice parameter is driven by the large magnitude of Bi distortions due to coupling of A-site displacement energy stabilization to the strain in the material~\cite{Cohen92p136,Halilov02p3443,Halilov04p174107,Ghita05p054114}. An increase of $c$ parameter also alleviates large direct and oxygen-mediated A-B repulsions created by the large Bi displacements~\cite{Grinberg02p909, Grinberg04p220101}.  
In contrast to other solid solutions, the B sites of BZT-PT are completely  occupied by ferroelectrically active (Zn and Ti) ions that can form short, covalent bonds with oxygen. For these cations, the large B-site displacement necessitated by large Bi displacements and A-B displacement coupling ~\cite{Grinberg02p909, Grinberg04p220101,Grinberg05p094111} are favorable.  
For these cations, there is no energetic penalty for making large off-center displacements necessitated by a combination of large Bi-cation displacements and oxygen bond order conservation.  
This stabilizes the tetragonal perovskite phase~\cite{Grinberg05p094111}.  While addition of BZT increases the $c$ axis parameter, the $a$ axis parameter is virtually unchanged with composition.  This is most likely due to the fact that Ti-O octahedra are rather stiff, and there is a significant energy cost for compressing Ti-O bonds along the short (010) and (001) directions~\cite{Halilov02p3443}.  The large ionic size of Zn$^{+2}$ also contributes to preventing decrease of the $a$ parameter.

The local structural distortions (reported above) for BZT-PT also help rationalize the dielectric properties.
Reported dielectric permittivity for the ($x$)BZT - (1-$x$)PT system~\cite{Suchomel05p262905} shows that increased BZT content causes a nonlinear rise in the temperature of the dielectric maximum ($T_{\epsilon,\rm max}$). The trend in $T_{\rm \epsilon,\rm max}$  can be understood by considering the changes of polarization magnitude with BZT content in conjunction with the previously established relationship that $T_{\epsilon,\rm max}$ is proportional to $P^2$ for PT-based solutions~\cite{Abrahams68p551,Juhas04p214101,Grinberg04p220101}.  Using Berry-phase calculated $P$ values for pure PT ($x$=0.0), $x$=0.25 and $x$=0.5 compositions, we see that the increase in $T_{\epsilon,\rm max}$ correlates with the enhancement of $P$, with extremely high $P$ values of 1.03$\pm0.05$~C/m$^2$  and 1.27$\pm$0.05~C/m$^2$ for the $x$=0.25 and $x$=0.50 compositions respectively (Table I). This is due to the large cation displacements present in these compositions.

The increase of $T_{\epsilon,\rm max}$ in BZT-PT is accompanied by a greater dispersion of the phase transition~\cite{Suchomel05p262905}.  It is unlikely that the appearance of diffuseness in the phase transition is due to the changes in local structure.  Analysis of relaxed DFT coordinates shows only small ($\approx$15$^{\circ}$) scatter in directions of cation displacements.  Such small scatter indicates that all the cations experience nearly the same local potential energy surface; this is typical of highly tetragonal PT-rich solid solutions~\cite{Grinberg04p220101,Egami01p33,Egami00p16} that do not exhibit diffusive or relaxor dielectric behavior~\cite{Landolt}.  We therefore conclude that the diffuse phase transitions observed in BZT-PT are not due to intrinsic properties but must be due to extrinsic effects.

Information about the microstructure of ceramic compositions in the ($x$)BZT - (1-$x$)PT system can also be extracted from the shape of the diffraction peaks in the experimental NPD data. In Figure 2a, we present the the 40-60$^\circ$ 2$\theta$ neutron diffraction data for the $x$=0.35 composition~\footnote[2]{Diffraction data for the full 2$\theta$ range and a complete summary of Rietveld refinement results are available in supplementary online material.}.  Arrows point out clearly visible asymmetric peak broadening on the high 2$\theta$ angle side of the $l$-containing peak reflections ((001), (102), etc..).  Although less marked, asymmetric broadening is also present on the low 2$\theta$ angle side of $h/k$-containing peak reflections. It is well known that peak broadening in diffraction experiments can have various causes, including small crystalline particle size or an internally stressed sample. However, both of these possible origins would result in symmetrical peak broadening.  In contrast, the broadening observed in the NPD presented here, and in XRD data of the ($x$)BZT - (1-$x$)PT system~\cite{Suchomel05p262905}, is quite strongly asymmetrical.

We propose that this unusual broadening may be explained by a microstructural model  based on local suppression of polarization and tetragonality near ferroelectric domain boundaries in highly tetragonal compositions with large $P$. In non-poled polycrystalline samples with randomly oriented domain polarizations, large elastic strain conditions make it energetically unfavorable for domains with non-aligned polarizations to lie adjacent to one another across a domain boundary. However, the strain can be minimized by the creation of a narrow region of locally suppressed polarization and tetragonality near the domain boundary, illustrated schematically in Figure 2b. In the strained near-boundary region, the $a$ and $c$ lattice parameters are distorted, with nearly cubic unit cells at the domain boundary.  When analyzed by diffraction, a sample possessing the microstructure shown in Fig. 2b will exhibit multiple sets of peak reflections.
In addition to the intrinsic reflections resulting from material in the center of the domains, the $l$-containing peak reflections of a sample with this structure will also exhibit a broad asymmetrical tail on the high 2$\theta$ angle side of the fundamental peak. This broadening effect is due to reflection contributions from reduced $c$ axes of the unit cells near the domain walls.  The increase in the $a$ and $b$ lattice parameters in the near-boundary region driven by conservation of unit cell volume  and rotation of polarization~\cite{Meyer02p104111} will give additional asymmetrical tails on the low 2$\theta$ angle side of the fundamental $h/k$-containing peak reflections.

To approximate the strained structure with continuously varying lattice parameters near the ferroelectric domain boundaries in compositions of the ($x$)BZT - (1-$x$)PT system, we used a two-phase model  in the Rietveld refinement of the neutron diffraction data for the x=0.20 and x=0.35 compositions.   Full details of this refinement are available online.  The model consists of high and low $c/a$ ratio tetragonal phases.  The high $c/a$ phase represents the undistorted intrinsic structure at the center of domains, while the low $c/a$ phase represents an average of the unit cells in the region of reduced polarization near the domain boundaries.   The estimated volume fractions of high and low $c/a$ phases are 0.90 and 0.10 respectively.  As shown in Figure 2b,  the Rietveld refinement results using this model agree with the experimental data.   The ionic displacements obtained by the Rietveld refinement for the high $c/a$ phase agree with our DFT calculated values (Table I).  Unsurprisingly, the low $c/a$ ratio values ($\approx$0.5~\AA\ for the A-site displacement and $\approx$0.25~\AA\ for the B-site displacement for $x$=0.20 and $x$=0.35) are quite different from DFT results, since the strain gradients present in the proposed microstructure were not included in the DFT calculations.

We believe that this non-uniform local environment in the ($x$)BZT - (1-$x$)PT system, rather than any intrinsic distortions, is the cause of the increased diffuseness of the dielectric transition observed for compositions with greater BZT content.
Local polarization  varies between the center of each 90$^\circ$ domain and the domain boundary.  This inhomogeneity leads to a distribution of local transition temperatures and a diffuse dielectric constant peak~\cite{Smolenskii61p2584}.  Additionally, the presence of uncompensated charge at the  90$^\circ$ domain boundaries gives rise to electric fields in the material~\cite{Meyer02p104111}.  It is likely that these also  contribute to the smearing of the phase transition.  Comparison of low and high $c/a$ displacements shows that as BZT content is increased, the differences between the domain interior and boundaries become more pronounced. It is this effect that leads to increased transition diffuseness.

In conclusion, we have examined the enhanced tetragonality  BZT-PT solid solution with a combination of experimental and first-principles methods.  Favorable coupling between A-site and B-site cation displacements gives rise to large polarization magnitudes.  Analysis of neutron diffraction data reveals that the high $c/a$ ratio and $P$ magnitude lead to significant changes in local structure near the 90$^\circ$ domain boundaries.  In turn, the inhomogeneity of the local structure smears out of the ferroelectric to parelectric phase transition.

This work was supported by the Office of Naval Research, under grant numbers N-000014-00-1-0372 and N00014-01-1-0860 and through the Center for Piezoelectrics by Design. We also acknowledge the support of the National Science Foundation, through the MRSEC program, grant No. DMR05-20020 and the use of  NPDF spectrometer at LANSCE, Los Alamos National Lab.   Computational support was provided by the Center for Piezoelectrics by Design, the  DoD HPCMO, DURIP and by the NSF CRIF program, Grant CHE-0131132.  

\bibliography{apssamp}


\appendix


\clearpage
\begin{table}[!t]
\caption{BZT-PT structural data obtained from DFT relaxed coordinates for $x$=0.25 and $x$=0.5 compositions and from Rietveld refinement of experimental neutron difraction data for $x$=0.20 and $x$=0.35 compositions. Cation displacements from center of oxygen cage in \AA.  Cation displacement scatter away from (100) direction ($\theta$) in degrees.  For experimentally investigated compositions, only average characteristics of A- and B-sites are available. Data are for the intrisic high $c/a$ structure, unless noted otherwise.  Low $c/a$ phase values are for the region next to domain boundary.  Polarization values in C/m$^2$.}

\begin{tabular}{l|cccc|ccc|cc}

\hline
&\multicolumn{1}{c}{}
&\multicolumn{1}{c}{Pb}
&\multicolumn{1}{c}{Bi}
&\multicolumn{1}{c}{Avg A}
&\multicolumn{1}{c}{Zn}
&\multicolumn{1}{c}{Ti}
&\multicolumn{1}{c}{Avg B}
&\multicolumn{1}{c}{}
&\multicolumn{1}{c}{}
\\

&\multicolumn{1}{c}{Data}
&\multicolumn{1}{c}{disp}
&\multicolumn{1}{c}{disp}
&\multicolumn{1}{c}{disp}
&\multicolumn{1}{c}{disp}
&\multicolumn{1}{c}{disp}
&\multicolumn{1}{c}{disp}
&\multicolumn{1}{c}{  $\theta_{\rm Pb}$}
&\multicolumn{1}{c}{$P$}
\\

\hline
\hline
0     &DFT& 0.44&  n/a &   0.44&  n/a  &  0.28&    0.28&  0 &   0.88\\
0.20  &NPD& --  &  --  &   0.56&  --   &  --  &    0.23&  --&   --  \\
0.25  &DFT& 0.49&  0.82&   0.59&  0.38 &  0.32&    0.33&  13&   1.03\\
0.35  &NPD& --  &  --  &   0.60&  --   &  --  &    0.29&  --&   --  \\
0.50  &DFT& 0.56&  0.90&   0.73&  0.50 &  0.34&    0.38&  18&   1.27\\

\end{tabular}
\end{table}

\clearpage
\begin{figure}
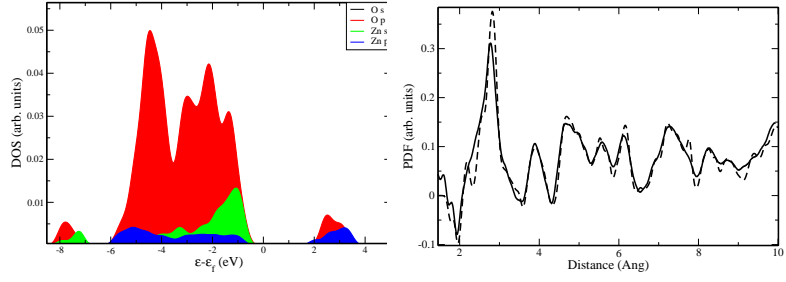

\centering

\includegraphics[width=2.0in]{ZnLDOS.eps}
\includegraphics[width=2.0in]{BZTpdfs2.eps}

\caption{{  
(a) Local density of states for Zn and O in $x$=0.25 BZT-PT. A cutoff radius of 2 a.u. was used to perform th\
e projection on atomic orbitals of Zn and O. Strong hybridization is present between  Zn 4$p$ and O 2$p$ orbitals.    This allows formation of short covalent Zn-O bonds and favors large Zn off-centering.
(b) ($x$)BZT-(1-$x$)PT pair distribution functions (PDF) obtained by neutron-scattering for $x$=0.20 (solid) and computed from relaxed DFT structures for $x$=0.25(dashed).  The experimental and theoretical data show excellent agreement.
 }}
\end{figure}

\clearpage
\begin{figure}
\centering

\includegraphics[width=2.3in]{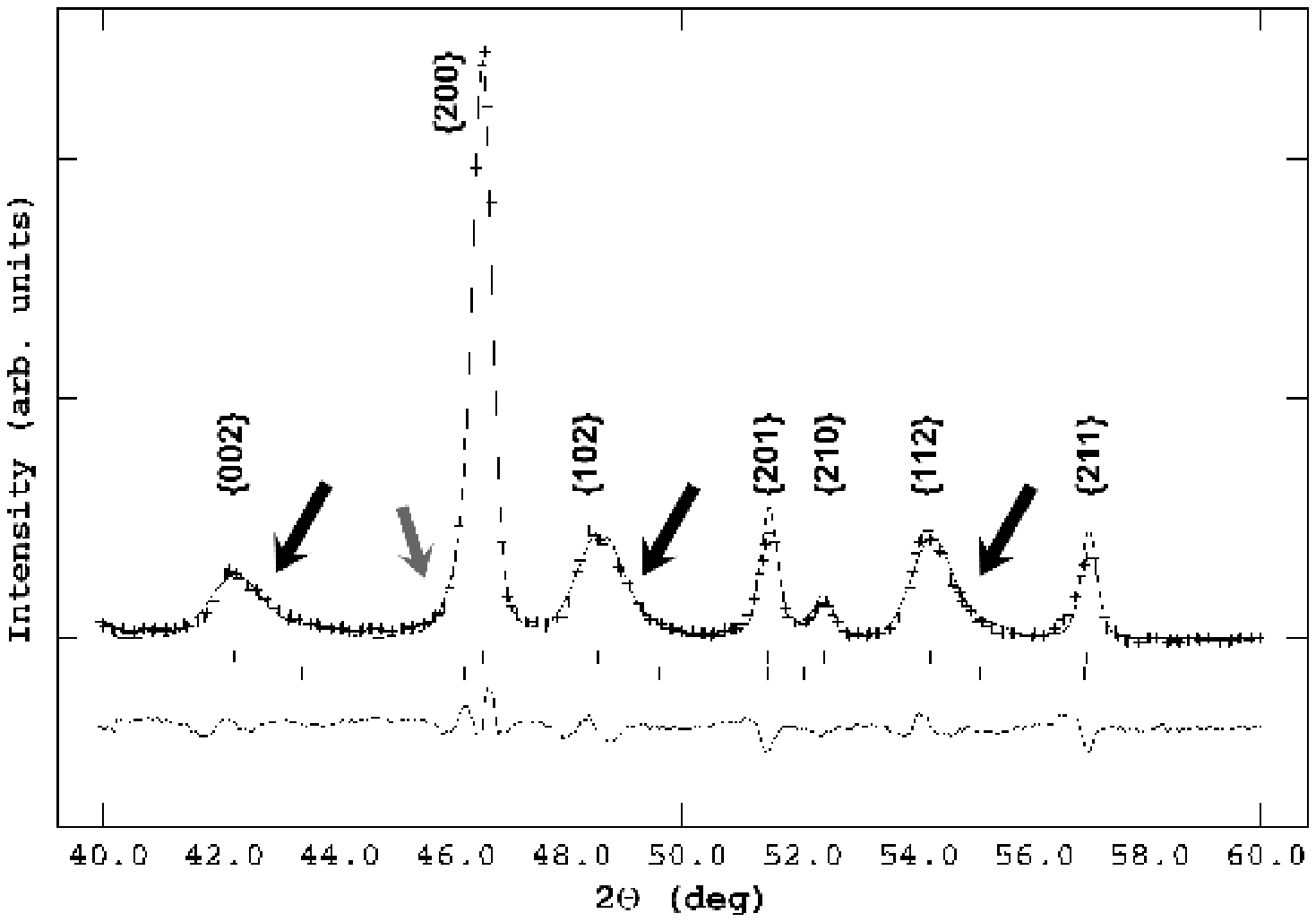}
\includegraphics[width=2.3in]{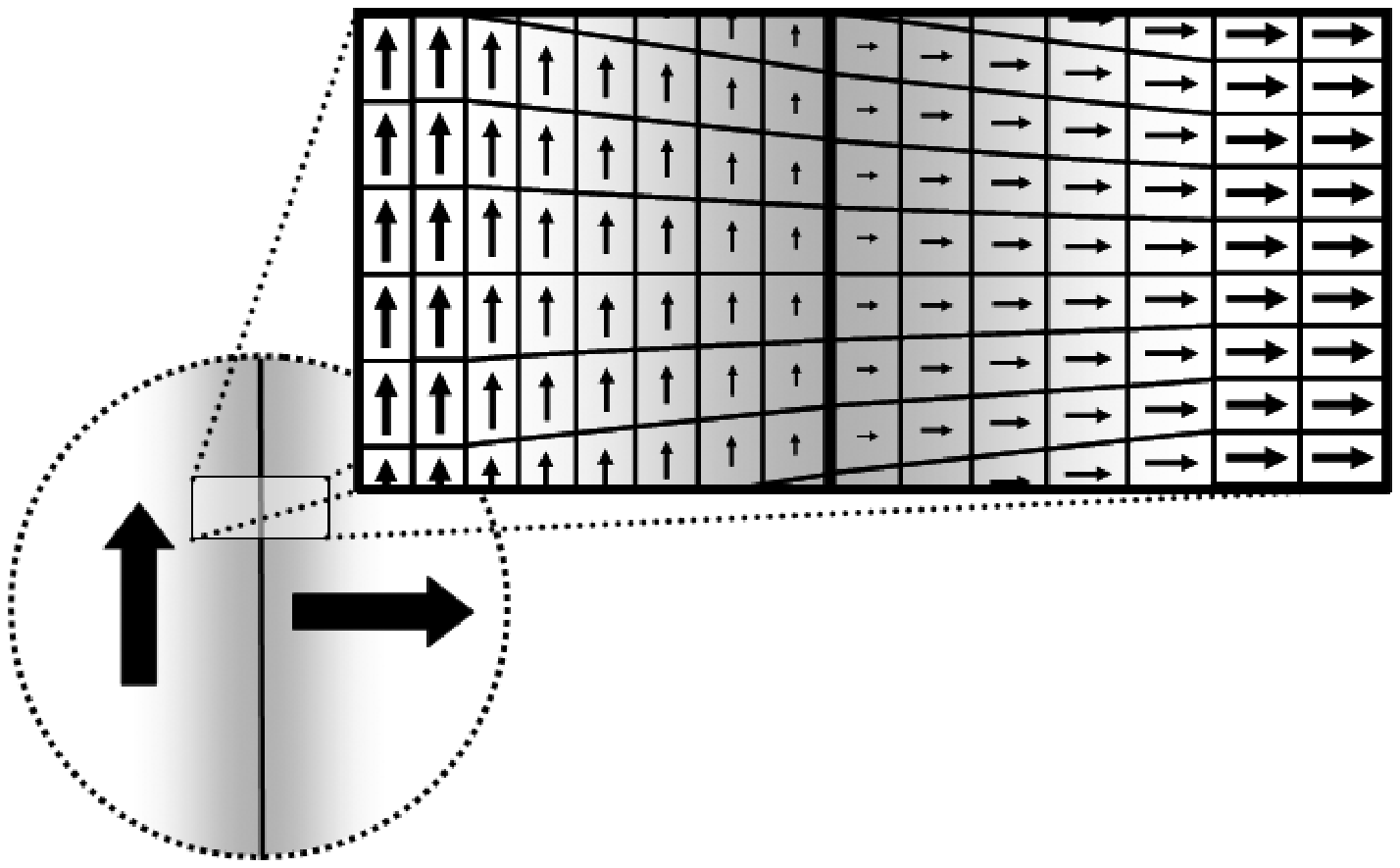}

\caption{{  
(a) Experimental and refined neutron diffraction data plotted in the 40-60$^\circ$ 2$\theta$ range  for the $x$=0.35 composition.  Crosses indicate experimental data, two-phase fit shown by solid line and fit difference is marked at bottom.  Black arrows point to asymmetric broadening of $l$-containing peaks, while the gray arrow highlights the more subtle $h,k$ peak broadening.
(b) Schematic illustration of locally suppressed tetragonality and polarization in the region near a 90$^\circ$ domain boundary.   Shaded area indicates the strained boundary region with suppressed polarization and tetragonality.
 }}
\end{figure}

\end{document}